
\input jytex.tex   
\typesize=10pt
\magnification=1200
\baselineskip=17truept
\hsize=6truein\vsize=8.5truein
\sectionnumstyle{blank}
\chapternumstyle{blank}
\chapternum=1
\sectionnum=1
\pagenum=0

\def\begintitle{\pagenumstyle{blank}\parindent=0pt\begin{narrow}[0.4in]}
\def\endtitle{\end{narrow}\newpage\pagenumstyle{arabic}}


\def\beginexercise{\vskip 20truept\parindent=0pt\begin{narrow}[10 truept]}
\def\endexercise{\vskip 10truept\end{narrow}}


\def\eql#1{\eqno\eqnlabel{#1}}
\def\ref{\reference}
\def\peq{\puteqn}
\def\pref{\putref}

\def\mgn{\marginnote}
\def\bex{\begin{exercise}}
\def\eex{\end{exercise}}

\font\open=msbm10 

\def\mbox#1{{\leavevmode\hbox{#1}}}
\def\hspace#1{{\phantom{\mbox#1}}}
\def\oZ{\mbox{\open\char90}}

\def\rS{{\rm S}}

\def\bka{{\bmit\kappa}}
\def\be{\beta}

\def\Ga{\Gamma}

\def\la{\lambda}

\def\om{\omega}

\def\si{\sigma}

\def\De{\Delta}


\def\frac#1/#2{\leavevmode\kern.1em
\raise.5ex\hbox{\the\scriptfont0 #1}\kern-.1em/\kern-.15em
\lower.25ex\hbox{\the\scriptfont0 #2}}
\def\sfrac#1/#2{\leavevmode\kern.1em
\raise.5ex\hbox{\the\scriptscriptfont0 #1}\kern-.1em/\kern-.15em
\lower.25ex\hbox{\the\scriptscriptfont0 #2}}

\def\gtorder{\mathrel{\raise.3ex\hbox{$>$}\mkern-14mu
             \lower0.6ex\hbox{$\sim$}}}
\def\ltorder{\mathrel{\raise.3ex\hbox{$<$}|mkern-14mu
             \lower0.6ex\hbox{\sim$}}}

\def\semidirprod{\rlap{\ss C}\raise1pt\hbox{$\mkern.75mu\times$}}
\def\for{\lower6pt\hbox{$\Big|$}}
\def\fish{\kern-.25em{\phantom{abcde}\over \phantom{abcde}}\kern-.25em}


\def\boxit#1{\vbox{\hrule\hbox{\vrule\kern3pt
        \vbox{\kern3pt#1\kern3pt}\kern3pt\vrule}\hrule}}
\def\dalemb#1#2{{\vbox{\hrule height .#2pt
        \hbox{\vrule width.#2pt height#1pt \kern#1pt
                \vrule width.#2pt}
        \hrule height.#2pt}}}

\def\ds{{|\!|}}        
\def\cd#1{{}_{\ds #1}} 

\def\noin{\noindent}


\def\cf{{\it cf }}
\def\pa{\partial}


\def\Tr{{\rm Tr\,}}

\def\3j#1#2#3#4#5#6{\left\lgroup\matrix{#1&#2&#3\cr#4&#5&#6\cr}
\right\rgroup}

\def\man{{\cal M}}

\def\m?{\mgn{?}}


\def\aop#1#2#3{{\it Ann. Phys.} {\bf {#1}} (19{#2}) #3}

\def\cqg#1#2#3{{\it Class. Quant. Grav.} {\bf {#1}} (19{#2}) #3}

\def\jpa#1#2#3{{\it J. Phys.} {\bf A{#1}} (19{#2}) #3}

\def\np#1#2#3{{\it Nucl. Phys.} {\bf B{#1}} (19{#2}) #3}

\def\pr#1#2#3{{\it Phys. Rev.} {\bf {#1}} (19{#2}) #3}

\def\dmj#1#2#3{{\it Duke Math. J.} {\bf {#1}} (19{#2}) #3}

\def\jdg#1#2#3{{\it J. Diff. Geom.} {\bf {#1}} (19{#2}) #3}

\def\ma#1#2#3{{\it Math. Ann.} {\bf {#1}} ({#2}) #3}


\begin{title}
\vglue 20truept
\righttext {MUTP/94/09}
\righttext{hep-th/9406002}
\leftline{\today}
\vskip 100truept
\centertext {\Bigfonts \bf Heat kernels on curved cones}
\vskip 15truept
\centertext{J.S.Dowker\footnote{Dowker@v2.ph.man.ac.uk}}
\vskip 7truept
\centertext{\it Department of Theoretical Physics,\\
The University of Manchester, Manchester, England.}
\vskip 60truept
\centertext {Abstract}
\begin{narrow}
A functorial derivation is presented of a scalar heat-kernel expansion
coefficient on
a manifold with a singular fixed point set of codimension two. Conformal
transformations are considered and the relevance of
an evanescent extrinsic curvature term is pointed out.
\end{narrow}
\vskip 5truept
\righttext {June 1994}
\vskip 100truept
\righttext{Typeset in \jyTeX}
\vfil
\end{title}
\pagenum=0

In an interesting article, Fursaev [\pref{Fur2}] has employed the general
technique of Donnelly [\pref{Donnelly}] to evaluate a particular heat-kernel
expansion coefficient for a manifold with a singular
fixed-point set of codimension two, the relevant underlying symmetry group
therefore being O(2) (with Killing vector $\pa\over\pa\phi$). The singularity
at a fixed point is a simple conical one in the sense that, near a point of
the fixed-point set, which we denote by ${\cal N}$, the manifold $\man_\be$
takes the product form ${\cal C}_\be\times{\cal N}$ where ${\cal C}_\be$ is a
cone of angle $\be$. When $\be=2\pi$ there is no singularity and
$\man_{2\pi}\equiv\man$ is smooth. Away from ${\cal N}$, $\man_\be$ and
$\man$ are locally identical.
As a totally geodesic submanifold of $\man$, ${\cal N}$ has vanishing
extrinsic curvatures.

The object of this letter is to present a simpler derivation of Fursaev's
expression and to draw attention to the consequences of conformal invariance.

The integrated heat-kernel expansion is written
$$K(t)\approx{1\over(4\pi t)^{d/2}}\sum_{n=0,1/2,\ldots}^\infty C_nt^n =
{1\over(4\pi t)^{d/2}}\sum_{n=0,1/2,\ldots}^\infty (A_n+D_n)t^n
\eql{hker}$$ where $A_n$ is the usual volume integral, over $\man_\be$,
of a local, scalar density involving
the curvature of $\man$. $D_n$ is a generalised boundary term. For simplicity
we assume that $\man$, and therefore $\man_\be$, does not have a conventional
(codimension one) boundary. The conical singularity is then responsible
for $D_n$.

The easier case when $\man$ is flat has been treated earlier and some special
curved situations have also been examined
[\pref{BandH,Dowkerccs,FandM,ChandD}].

Both $A_n$ and $D_n$ depend on the conical angle $\be$, the first quantity in
a trivial fashion. Because of the O(2) symmetry, there is the simple volume
relation $A_n(\be)=(\be/2\pi)A_n(2\pi)$ which serves to fix $A_n(\be)$.

Fursaev writes, following Donnelly, the general structure of the $D_n$ as
$$D_n=\sum_{k=1}^n \be P_k(\be)G_{nk}
\eql{genst}$$
where the $G_{nk}$ are volume integrals over ${\cal N}$ of a local
expression involving the curvature (of $\man$) its covariant derivatives and
the two normals associated with ${\cal N}$.
The $P_k(\be)$ are certain polynomials in $B\equiv2\pi/\be$. They are related
to generalised Bernoulli polynomials and each has the factor $(B^2-1)$.

We prefer to write in general
$$D_n={2\pi\over B}\sum_k (B^{2k}-1)H_{nk}.
\eql{gencoeff}$$

Instead of using more differential geometry, one can proceed by
writing down the general form of the $H_{nk}$ and finding sufficient
special cases to fix the unknown numerical parameters. We will do this for the
$D_2$ coefficient and will reproduce Fursaev's formula.

Simple arguments show that the integrand, $h_{nk}$, of $H_{nk}$ has dimension
$L^{2-2n}$ and so its most general form is
$$h_{2k}=L_kR+M_kR_{\mu\nu}{\bf n}^\nu{\bf .n}^\mu+N_kR_{\mu\nu\rho\si}
({\bf n}^{\mu}{\bf .n}^{\rho})({\bf n}^{\nu}{\bf .n}^{\si})
\eql{h21}$$ where the $L,M,N$ are independent of the dimension $d$, so long
as the heat operator is.

Regarding notation, $n^\mu_i$ $(i=1$ to codimension, = 2 here) are the inward
normals to ${\cal N}$. The indices $i,j$ refer to the normal fibre, $\mu,\nu$
to the whole manifold, $\man_\be$, and $a,b$ intrinsically to ${\cal N}$.

There are no covariant derivatives in (\peq{h21}) so, in order to find the
$L,M$ and $N$, it
is enough to consider the special case of the cyclically factored $d$-sphere,
$\rS^d/\oZ_q$. Then $B=q\in\oZ$ and (\peq{h21}) becomes\mgn{MTW Convensions}
$$h_{2k}=L_k d(d-1)+2M_k(d-1)+2N_k
\eql{h22}$$ where we have used the fact that the normals $n^\mu_i$ are mutually
orthonormal. The fixed-point set is a $(d-2)$-sphere and so the integrated
expression is
$$H_{2k}= |\rS^{d-2}|(L_k d(d-1)+2M_k(d-1)+2N_k).
\eql{inth2}$$
Our curvature conventions are those of Hawking and Ellis [\pref{HandE}].

The coefficients in the scalar heat kernel expansion for non-minimal coupling,
$\xi R$, have been determined in Chang and Dowker [\pref{ChandD}] when $\xi$
equals $(d-1)/4d$. Since $R$ is constant,
it is easy to find the coefficients in the general case. An equivalent
expression for the minimal value, $\xi=0$, follows from our previous work
[\pref{Dow}].

Excluding the $A_2$ part from the complete coefficient, we find
$$\eqalign{
D_2&={(q^2-1)\over360q}\bigg({(4\pi)^{d/2}\Ga({d-2\over2}\over\Ga(d-2)}\bigg)
\bigg(\!15(d-1)\big(d-1-4\xi d\big)-(d-3)(2q^2-3+5d)\!\bigg)\cr
\noalign{\vskip 4truept}
&={2\pi\over q}{(q^2-1)\over360}|\rS^{d-2}|
\bigg(10d(d-1)\big(1-6\xi\big)-2(d-3)(q^2+1)\bigg).\cr}
\eql{dee21}$$
{}From (\peq{gencoeff})
$$\eqalign{
D_2&={2\pi\over q}(q^2-1) \big(H_{2,1}+H_{2,2}(q^2+1)\big)\cr
&={2\pi\over q}(q^2-1)|\rS^{d-2}| \bigg(L_1 d(d-1)+2M_1(d-1)+2N_1\cr
&\hspace{*****}+\big(L_2 d(d-1)+2M_2(d-1)+2N_2\big)\big(q^2+1\big)\bigg).
\cr}\eql{dee22}$$

Equating (\peq{dee21}) and (\peq{dee22}) we find $L_1=(1-6\xi)/72$,
$M_1=N_1=0$ and $L_2=0$, $N_2=-2M_2=1/360$. These values agree with those of
Fursaev [\pref{Fur2}].

It is a general theorem [\pref{DandK,KCD}] that in $d$-dimensions,
if $\xi=(d-2)/4(d-1)$, the
total heat-kernel coefficient $C^{(d)}_{d/2}$ is invariant under Weyl
rescalings.
It is useful to check this for the present situation.

In four dimensions, setting $\xi=1/6$, we have

$$D_2={\pi\over360 B}(B^4-1)\int_{\cal N}h^{1/2}d^{d-2}y\,
\big(2R_{\mu\nu\rho\si}({\bf n}^{\mu}{\bf .n}^{\rho})({\bf n}^{\nu}{\bf .n}
^{\si})-
R_{\mu\nu}{\bf n}^\nu{\bf .n}^\mu\big)
\eql{confdee2}$$
where $y$ are the coordinates and $h_{ab}$ the induced metric on ${\cal N}$.
The curvatures are those of the embedding space $\man$, evaluated on
${\cal N}$.

Under the metric rescaling $g_{\mu\nu}\to e^{-2\om}g_{\mu\nu}$ we have
${\bf n}^\mu\to e^\om {\bf n}^\mu$ and $h_{ab}\to e^{-2\om}h_{ab}$, $h\to
e^{2(2-d)\om}h$.

It is assumed that there are no singularities in the scaling so that angles, in
particular $\be$, are preserved.

Since $A_2$ is, up to a constant factor, the conventional, boundaryless
coefficient, it will be conformally invariant and
the total invariance should follow from that of $D_2$.

Standard formulae give\mgn{MTW. R.... same as Eisenhart but R.. opposite.
$\si=-\om$}
$$\eqalign{
&R_{\mu\nu}{\bf n}^\nu{\bf .n}^\mu\to
e^{2\om}\big(R_{\mu\nu}{\bf n}^\nu{\bf .n}^\mu+2\om_{\mu\nu}{\bf n}^\mu
{\bf .n}^\nu+
2\De_2\om+4\De_1\om\big)\cr
\noalign{\vskip 4truept}
&R_{\mu\nu\rho\si}({\bf n}^{\mu}{\bf .n}^{\rho})({\bf n}^{\nu}{\bf .n}^{\si})
\to e^{2\om}\big(R_{\mu\nu\rho\si}({\bf n}^{\mu}{\bf .n}^{\rho})
({\bf n}^{\nu}{\bf .n}^{\si})+
2\om_{\mu\nu}{\bf n}^\mu{\bf .n}^\nu+2\De_1\om\big)\cr}
$$where $\om_{\mu\nu}=\om\cd{\mu\nu}+\om\cd\mu\om\cd\nu$.

Looking at (\peq{confdee2}), we can see immediately that the $\De_1\om$ terms
cancel and the conformal change in
$ h^{1/2}\big(2R_{\mu\nu\rho\si}({\bf n}^{\mu}{\bf .n}^{\rho})
({\bf n}^{\nu}{\bf .n}^{\si})-R_{\mu\nu}{\bf n}^\nu{\bf .n}^\mu\big)$
equals
$$ h^{1/2}\big(2\om_{\mu\nu}{\bf n}^\mu{\bf .n}^\nu-2\De_2\om\big)=
h^{1/2}\big(
2\om\cd{\mu\nu}{\bf n}^\mu{\bf .n}^\nu+2\om\cd\mu\om\cd\nu
{\bf n}^\mu{\bf .n}^\nu-2\De_2\om\big).
\eql{confch}$$

 We now need the relation between the embedding and intrinsic
Laplacians evaluated on a submanifold (of any codimension),
$$\De_2\om+\bka{\bf .n}^\mu\om\cd\mu=\widehat\De_2\om
+{\bf n}^\mu{\bf .n}^\nu\om\cd{\mu\nu},
$$ where $\bka$ is the trace of the extrinsic curvature, $\bka^{ab}$ of the
submanifold.

The change (\peq{confch}) can then be written
$$ h^{1/2}\big(2\om\cd\mu\om\cd\nu {\bf n}^\mu{\bf.n}^\nu+2
\bka{\bf.n}^\mu\om\cd\mu-2\widehat\De_2\om\big),
$$which should vanish.
Indeed the intrinsic Laplacian, $\widehat\De_2\om$,
does integrate to zero, but there seems to be a piece left over.

The solution to this dilemma is that a term proportional to $\bka.\bka$
should be included in the general form (\peq{h21}), as allowed by
dimensions. This term vanishes for fixed-point sets but must be retained for
consistency under conformal transformations.

Conformal invariance tells us that there is no such contribution to
$h_{2,1}$ but that  $h_{2,2}$ should be augmented by a term
$$-{1\over2}\bka{\bf.}\bka+\la\big(\bka{\bf.}\bka-
2\Tr(\bka{\bf.}\bka)\big).
\eql{extra}$$

This can be checked from the conformal transformation
$\bka^{ab}\to e^{-\om}\big(\bka_{ab}+h_{ab}{\bf n}^\mu\om\cd\mu\big),
$ whence $\bka\to e^{\om}\big(\bka+(d-2){\bf n}^\mu\om\cd\mu\big)
$ for codimension two.

The parameter $\la$ remains undetermined at the moment because the quantity in
brackets in (\peq{extra}) is conformally invariant in two dimensions.

These general considerations regarding the heat-kernel can be extended to the
higher coefficients and to
submanifolds of other codimensions, \cf Cheeger [\pref{Cheeger}].

\vskip 2truept
\noin{\bf{References}}
\vskip 1truept
\begin{putreferences}
\ref{Donnelly}{H.Donnelly \ma{224}{1976}161.}
\ref{Dow}{J.S.Dowker {\sl Effective actions on spherical domains},
{\it Comm.Math.Phys}, in the press.}
\ref{KCD}{G.Kennedy, R.Critchley and J.S.Dowker \aop{125}{80}{346}.}
\ref{Fur2}{D.V.Fursaev {\sl Spectral geometry and one-loop divergences on
manifolds with conical singularities}, JINR preprint DSF-13/94,
hep-th/9405143.}
\ref{HandE}{S.W.Hawking and G.F.R.Ellis {\sl The large scale structure of
space-time} Cambridge University Press, 1973.}
\ref{DandK}{J.S.Dowker and G.Kennedy \jpa{11}{78}{895}.}
\ref{ChandD}{Peter Chang and J.S.Dowker \np{395}{93}{407}.}
\ref{FandM}{D.V.Fursaev and G.Miele \pr{D49}{94}{987}.}
\ref{Dowkerccs}{J.S.Dowker \cqg{4}{87}{L157}.}
\ref{BandH}{J.Br\"uning and E.Heintze \dmj{51}{84}{959}.}
\ref{Cheeger}{J.Cheeger \jdg{18}{83}{575}.}
\end{putreferences}
\bye